\documentclass[twocolumn,showpacs,preprintnumbers,amsmath,amssymb,floatfix,prd,superscriptaddress]{revtex4}

\usepackage{graphicx}
\usepackage{dcolumn}
\usepackage{bm}
\usepackage{lineno}
\usepackage{setspace}
\usepackage{revsymb}
\usepackage{url}
\usepackage{epsfig}
\usepackage{color}

\newcommand{\met}{\mbox{$E_{T}\!\!\!\!\!\!\!/\,\,\,\,$}}
\newcommand{\cl}{\mbox{C.\ \!\!L.\ \!\!}}
\newcommand{\cll}{\mbox{C.\ \!\!L}}

\begin{document}

\title{\large \bf \boldmath Search for new physics in trilepton events and limits on the associated chargino-neutralino production at CDF}

\affiliation{Institute of Physics, Academia Sinica, Taipei, Taiwan 11529, Republic of China}
\affiliation{Argonne National Laboratory, Argonne, Illinois 60439, USA}
\affiliation{University of Athens, 157 71 Athens, Greece}
\affiliation{Institut de Fisica d'Altes Energies, ICREA, Universitat Autonoma de Barcelona, E-08193, Bellaterra (Barcelona), Spain}
\affiliation{Baylor University, Waco, Texas 76798, USA}
\affiliation{Istituto Nazionale di Fisica Nucleare Bologna, \ensuremath{^{ii}}University of Bologna, I-40127 Bologna, Italy}
\affiliation{University of California, Davis, Davis, California 95616, USA}
\affiliation{University of California, Los Angeles, Los Angeles, California 90024, USA}
\affiliation{Instituto de Fisica de Cantabria, CSIC-University of Cantabria, 39005 Santander, Spain}
\affiliation{Carnegie Mellon University, Pittsburgh, Pennsylvania 15213, USA}
\affiliation{Enrico Fermi Institute, University of Chicago, Chicago, Illinois 60637, USA}
\affiliation{Comenius University, 842 48 Bratislava, Slovakia; Institute of Experimental Physics, 040 01 Kosice, Slovakia}
\affiliation{Joint Institute for Nuclear Research, RU-141980 Dubna, Russia}
\affiliation{Duke University, Durham, North Carolina 27708, USA}
\affiliation{Fermi National Accelerator Laboratory, Batavia, Illinois 60510, USA}
\affiliation{University of Florida, Gainesville, Florida 32611, USA}
\affiliation{Laboratori Nazionali di Frascati, Istituto Nazionale di Fisica Nucleare, I-00044 Frascati, Italy}
\affiliation{University of Geneva, CH-1211 Geneva 4, Switzerland}
\affiliation{Glasgow University, Glasgow G12 8QQ, United Kingdom}
\affiliation{Harvard University, Cambridge, Massachusetts 02138, USA}
\affiliation{Division of High Energy Physics, Department of Physics, University of Helsinki, FIN-00014, Helsinki, Finland; Helsinki Institute of Physics, FIN-00014, Helsinki, Finland}
\affiliation{University of Illinois, Urbana, Illinois 61801, USA}
\affiliation{The Johns Hopkins University, Baltimore, Maryland 21218, USA}
\affiliation{Institut f\"{u}r Experimentelle Kernphysik, Karlsruhe Institute of Technology, D-76131 Karlsruhe, Germany}
\affiliation{Center for High Energy Physics: Kyungpook National University, Daegu 702-701, Korea; Seoul National University, Seoul 151-742, Korea; Sungkyunkwan University, Suwon 440-746, Korea; Korea Institute of Science and Technology Information, Daejeon 305-806, Korea; Chonnam National University, Gwangju 500-757, Korea; Chonbuk National University, Jeonju 561-756, Korea; Ewha Womans University, Seoul, 120-750, Korea}
\affiliation{Ernest Orlando Lawrence Berkeley National Laboratory, Berkeley, California 94720, USA}
\affiliation{University of Liverpool, Liverpool L69 7ZE, United Kingdom}
\affiliation{University College London, London WC1E 6BT, United Kingdom}
\affiliation{Centro de Investigaciones Energeticas Medioambientales y Tecnologicas, E-28040 Madrid, Spain}
\affiliation{Massachusetts Institute of Technology, Cambridge, Massachusetts 02139, USA}
\affiliation{University of Michigan, Ann Arbor, Michigan 48109, USA}
\affiliation{Michigan State University, East Lansing, Michigan 48824, USA}
\affiliation{Institution for Theoretical and Experimental Physics, ITEP, Moscow 117259, Russia}
\affiliation{University of New Mexico, Albuquerque, New Mexico 87131, USA}
\affiliation{The Ohio State University, Columbus, Ohio 43210, USA}
\affiliation{Okayama University, Okayama 700-8530, Japan}
\affiliation{Osaka City University, Osaka 558-8585, Japan}
\affiliation{University of Oxford, Oxford OX1 3RH, United Kingdom}
\affiliation{Istituto Nazionale di Fisica Nucleare, Sezione di Padova, \ensuremath{^{jj}}University of Padova, I-35131 Padova, Italy}
\affiliation{University of Pennsylvania, Philadelphia, Pennsylvania 19104, USA}
\affiliation{Istituto Nazionale di Fisica Nucleare Pisa, \ensuremath{^{kk}}University of Pisa, \ensuremath{^{ll}}University of Siena, \ensuremath{^{mm}}Scuola Normale Superiore, I-56127 Pisa, Italy, \ensuremath{^{nn}}INFN Pavia, I-27100 Pavia, Italy, \ensuremath{^{oo}}University of Pavia, I-27100 Pavia, Italy}
\affiliation{University of Pittsburgh, Pittsburgh, Pennsylvania 15260, USA}
\affiliation{Purdue University, West Lafayette, Indiana 47907, USA}
\affiliation{University of Rochester, Rochester, New York 14627, USA}
\affiliation{The Rockefeller University, New York, New York 10065, USA}
\affiliation{Istituto Nazionale di Fisica Nucleare, Sezione di Roma 1, \ensuremath{^{pp}}Sapienza Universit\`{a} di Roma, I-00185 Roma, Italy}
\affiliation{Mitchell Institute for Fundamental Physics and Astronomy, Texas A\&M University, College Station, Texas 77843, USA}
\affiliation{Istituto Nazionale di Fisica Nucleare Trieste, \ensuremath{^{qq}}Gruppo Collegato di Udine, \ensuremath{^{rr}}University of Udine, I-33100 Udine, Italy, \ensuremath{^{ss}}University of Trieste, I-34127 Trieste, Italy}
\affiliation{University of Tsukuba, Tsukuba, Ibaraki 305, Japan}
\affiliation{Tufts University, Medford, Massachusetts 02155, USA}
\affiliation{University of Virginia, Charlottesville, Virginia 22906, USA}
\affiliation{Waseda University, Tokyo 169, Japan}
\affiliation{Wayne State University, Detroit, Michigan 48201, USA}
\affiliation{University of Wisconsin, Madison, Wisconsin 53706, USA}
\affiliation{Yale University, New Haven, Connecticut 06520, USA}

\author{T.~Aaltonen}
\affiliation{Division of High Energy Physics, Department of Physics, University of Helsinki, FIN-00014, Helsinki, Finland; Helsinki Institute of Physics, FIN-00014, Helsinki, Finland}
\author{S.~Amerio\ensuremath{^{jj}}}
\affiliation{Istituto Nazionale di Fisica Nucleare, Sezione di Padova, \ensuremath{^{jj}}University of Padova, I-35131 Padova, Italy}
\author{D.~Amidei}
\affiliation{University of Michigan, Ann Arbor, Michigan 48109, USA}
\author{A.~Anastassov\ensuremath{^{v}}}
\affiliation{Fermi National Accelerator Laboratory, Batavia, Illinois 60510, USA}
\author{A.~Annovi}
\affiliation{Laboratori Nazionali di Frascati, Istituto Nazionale di Fisica Nucleare, I-00044 Frascati, Italy}
\author{J.~Antos}
\affiliation{Comenius University, 842 48 Bratislava, Slovakia; Institute of Experimental Physics, 040 01 Kosice, Slovakia}
\author{G.~Apollinari}
\affiliation{Fermi National Accelerator Laboratory, Batavia, Illinois 60510, USA}
\author{J.A.~Appel}
\affiliation{Fermi National Accelerator Laboratory, Batavia, Illinois 60510, USA}
\author{T.~Arisawa}
\affiliation{Waseda University, Tokyo 169, Japan}
\author{A.~Artikov}
\affiliation{Joint Institute for Nuclear Research, RU-141980 Dubna, Russia}
% ...

\collaboration{CDF Collaboration}
\altaffiliation[With visitors from]{
\ensuremath{^{a}}University of British Columbia, Vancouver, BC V6T 1Z1, Canada,
\ensuremath{^{b}}Istituto Nazionale di Fisica Nucleare, Sezione di Cagliari, 09042 Monserrato (Cagliari), Italy,
\ensuremath{^{c}}University of California Irvine, Irvine, CA 92697, USA,
\ensuremath{^{d}}Institute of Physics, Academy of Sciences of the Czech Republic, 182~21, Czech Republic,
\ensuremath{^{e}}CERN, CH-1211 Geneva, Switzerland,
\ensuremath{^{f}}Cornell University, Ithaca, NY 14853, USA,
\ensuremath{^{g}}University of Cyprus, Nicosia CY-1678, Cyprus,
\ensuremath{^{h}}Office of Science, U.S. Department of Energy, Washington, DC 20585, USA,
\ensuremath{^{i}}University College Dublin, Dublin 4, Ireland,
\ensuremath{^{j}}ETH, 8092 Z\"{u}rich, Switzerland,
\ensuremath{^{k}}University of Fukui, Fukui City, Fukui Prefecture, Japan 910-0017,
\ensuremath{^{l}}Universidad Iberoamericana, Lomas de Santa Fe, M\'{e}xico, C.P. 01219, Distrito Federal,
\ensuremath{^{m}}University of Iowa, Iowa City, IA 52242, USA,
\ensuremath{^{n}}Kinki University, Higashi-Osaka City, Japan 577-8502,
\ensuremath{^{o}}Kansas State University, Manhattan, KS 66506, USA,
\ensuremath{^{p}}Brookhaven National Laboratory, Upton, NY 11973, USA,
\ensuremath{^{q}}Queen Mary, University of London, London, E1 4NS, United Kingdom,
\ensuremath{^{r}}University of Melbourne, Victoria 3010, Australia,
\ensuremath{^{s}}Muons, Inc., Batavia, IL 60510, USA,
\ensuremath{^{t}}Nagasaki Institute of Applied Science, Nagasaki 851-0193, Japan,
\ensuremath{^{u}}National Research Nuclear University, Moscow 115409, Russia,
\ensuremath{^{v}}Northwestern University, Evanston, IL 60208, USA,
\ensuremath{^{w}}University of Notre Dame, Notre Dame, IN 46556, USA,
\ensuremath{^{x}}Universidad de Oviedo, E-33007 Oviedo, Spain,
\ensuremath{^{y}}CNRS-IN2P3, Paris, F-75205 France,
\ensuremath{^{z}}Universidad Tecnica Federico Santa Maria, 110v Valparaiso, Chile,
\ensuremath{^{aa}}The University of Jordan, Amman 11942, Jordan,
\ensuremath{^{bb}}Universite catholique de Louvain, 1348 Louvain-La-Neuve, Belgium,
\ensuremath{^{cc}}University of Z\"{u}rich, 8006 Z\"{u}rich, Switzerland,
\ensuremath{^{dd}}Massachusetts General Hospital, Boston, MA 02114 USA,
\ensuremath{^{ee}}Harvard Medical School, Boston, MA 02114 USA,
\ensuremath{^{ff}}Hampton University, Hampton, VA 23668, USA,
\ensuremath{^{gg}}Los Alamos National Laboratory, Los Alamos, NM 87544, USA,
\ensuremath{^{hh}}Universit\`{a} degli Studi di Napoli Federico I, I-80138 Napoli, Italy
}
\noaffiliation

\date{September 28, 2013}

\begin{abstract}

We perform a search for new physics using final states consisting of three leptons and a large imbalance in transverse momentum resulting from proton-antiproton collisions at 1.96 TeV center-of-mass energy.
We use data corresponding to 5.8 fb$^{-1}$ of integrated luminosity recorded by the CDF II detector at the Tevatron collider.
Our main objective is to investigate possible new low-momentum (down to 5 GeV/$c$) multi-leptonic final states not investigated by LHC experiments.  Relative to previous CDF analyses, we expand the geometric and kinematic coverage of 
electrons and muons and utilize 
tau leptons that decay hadronically.  Inclusion of tau leptons is particularly important for supersymmetry (SUSY) searches.
The results are consistent with standard-model predictions.
By optimizing our event selection to increase sensitivity to the
minimal supergravity (mSUGRA) SUSY model, 
we set limits on the associated production of chargino and neutralino, the SUSY partners of the electroweak gauge bosons.  
We exclude cross sections up to $0.1$ pb and chargino masses up to $168$ GeV/$c^2$
at 95\% $\cl$, for a suited set of mSUGRA parameters.
We also exclude a region of the two-dimensional space of the masses of the neutralino and the 
supersymmetric partner of the tau lepton, not previously excluded at the Tevatron.
\end{abstract}

\maketitle

Although extremely successful, the standard model (SM) of particles and fields 
leaves many questions unanswered, including the origin 
of dark matter, the incorporation of gravity, and the hierarchy between the weak-interaction 
and Planck energy scales.  New physics that would address these issues could
be directly discovered in particle topologies that are characterized by low SM background.
Such topologies include final states involving three charged leptons 
({\it trilepton}) in hadron collisions.
A trilepton signal is predicted by several 
new-physics processes, including lepton-flavor-violating tau-lepton decays \cite{lfv}, heavy-neutrino decays in 
see-saw models \cite{seesaw}, Higgs-boson decays in inert doublet models \cite{inert},
Kaluza-Klein-graviton decays in low-scale warped-extra-dimension 
models \cite{kk}, and, most notably, in chargino and neutralino decays in supersymmetric (SUSY \cite{susy}) processes.

In this Letter, we present a blind, model-independent search for new physics in the trilepton plus missing (i.e., unbalanced) transverse momentum ($\met$) final state
at the Fermilab Tevatron collider, where protons and antiprotons collided with a center-of-mass energy of 1.96 TeV.
We illustrate the sensitivity of our search in a particular class of SUSY models, involving minimal supergravity (mSUGRA\cite{susybook}), with a small number of parameters \cite{msugra}.  One of the low-background processes for the discovery of SUSY particles 
in proton-antiproton collisions is the associated chargino-neutralino ($\tilde{\chi}^{\pm}_1 \tilde{\chi}^{0}_2$) production and the resulting 
trilepton+$\met$ final state: $p\bar{p} \rightarrow \tilde{\chi}^{\pm}_1 \tilde{\chi}^{0}_2$, followed by, e.g., $\tilde{\chi}^{\pm}_1\rightarrow \ell\nu\tilde{\chi}^{0}_1$ and $\tilde{\chi}^{0}_2\rightarrow\ell\ell\tilde{\chi}^{0}_1$ \cite{baer}.
 The lightest chargino $\tilde{\chi}^{\pm}_1$ and the next-to-lightest neutralino $\tilde{\chi}^{0}_2$ are supersymmetric partners of the gauge bosons, $\ell$ indicates an electron ($e$), a muon ($\mu$), or tau lepton ($\tau$), and ${\chi}^0_1$ is the lightest neutralino, assumed to be stable and escaping detection, and therefore contributing to the missing transverse momentum.  
After completing our model-independent search, we optimize our analysis specifically for the associated chargino-neutralino production.  

The CDF experiment has previously searched for this signature using data from up to 3.2 fb$^{-1}$ \cite{all,anadi,unm, rutgers, ucsd} of Run II integrated luminosity.  
The latest D0 trilepton analysis \cite{d0} used 2.3 fb$^{-1}$ of integrated luminosity.
The ATLAS collaboration has published a trilepton+$\met$ search using 2.3 fb$^{-1}$ \cite{atlas} and the CMS 
collaboration has published results using a luminosity of 5 fb$^{-1}$ \cite{cms,cms2}.
We present here an analysis with 5.8 fb$^{-1}$ of integrated luminosity.  
This search is significantly improved compared to the previous CDF trilepton searches.  
We expand the acceptance to cover the forward region of the detector for both electrons and muons, include 
(as third leptons) tau leptons decaying hadronically,
and allow lower momenta for our leptonic candidates (down to 5 GeV/$c$), within the 
constraints of the candidate identification and online event-selection ({\it trigger}) requirements.
Lower (and forward) leptonic momenta allow us to investigate in a model-independent way
either the direct decay of new light particles or the chain decay of particles with similar masses.   
The inclusion of tau leptons is also motivated by the high branching ratio of chargino and neutralino 
decays to the lightest supersymmetric lepton ($\tilde{\ell}$), 
typically the stau ($\tilde{\tau}$), which preferably decays to a tau lepton.

CDF II \cite{cdfDetector} is a multipurpose cylindrical detector with a projective-tower 
calorimeter geometry and an excellent lepton identification capability.  It operated at Fermilab's Tevatron collider.
In CDF's coordinate system, the positive $z$-axis is defined by the proton beam direction and the positive $y$-axis by the vertically-upward 
direction.  The detector is approximately symmetric in the $\eta$ and $\phi$ directions, where the pseudorapidity 
$\eta$ is defined as $\eta = -\ln[\tan (\theta/2)]$, $\theta$ is the polar angle with respect to the $z$-axis, and $\phi$ is the azimuthal angle.

The momentum $p$ of charged particles is measured with a tracking system composed of a seven-layer silicon
strip detector and a 96-layer drift chamber;
both are located inside a solenoid aligned along the beam axis and providing 
a magnetic field of 1.4~T.  
The tracking efficiency is nearly 100\% in the 
central region ($|\eta|<1$) and decreases in the forward region ($1<|\eta|<2.8$).  
Electrons can be identified in the forward region by using tracks reconstructed
using only silicon-tracker information.
Electromagnetic and hadronic calorimeters surround the solenoid and measure the energies of collision products up to $|\eta|=3.6$. 
Drift chambers and scintillators are installed outside
the hadronic calorimeter to detect muons with $|\eta|<1.4$. 
Gas Cherenkov counters \cite{cherenkov} downstream at small angles with respect to the two beams measure the average number of inelastic $p\bar{p}$ collisions 
per bunch crossing and thereby determine the collisions' luminosity.  A pipelined three-level trigger 
system \cite{trigger} that combines hardware and software is used for filtering 
the collision data.

We perform an analysis of dielectron+$\ell'$ and dimuon+$\ell'$ data collected 
with single high-transverse-momentum ($p_T\equiv p\sin\theta >18$~GeV/$c$) central electron and central muon triggers, respectively.  
The third object $\ell'$ can be 
an electron, a muon, a tau lepton, or an isolated track (isoTrack).
Events where the two highest-in-$p_T$ leptons are $e\mu$ or $\mu e$ 
are included only if the third object is an electron or muon.
No requirement is applied on the charge of the leptons.
To ensure a uniform trigger response,
we require a central electron or central muon with $p_T>20$~GeV/$c$.
The second and third electron or muon can be detected in 
either the central or the forward region of the detector and is required to have $p_T>5$~GeV/$c$.  
The tracking system provides the 
direction of the electrons, whereas the magnitude of their momentum is
determined from the energy deposited in the calorimeters. 
This energy deposition is required to match the track geometrically, and 
the lateral shape of the deposition must be consistent with that expected 
for electrons.   For the muons, both
direction and energy are determined from a track that is 
matched with signals from the muon detectors.
The additional transverse energy deposited in the calorimeter
in a cone of $\Delta R = \sqrt{(\Delta\phi)^2+(\Delta\eta)^2}=0.4$
around each electron or muon must be less than 10\% of the lepton's transverse energy, if the lepton has $p_T>20$~GeV/$c$.  Otherwise, we require that this additional energy is less than 2 GeV.
The electrons and muons are required to be separated by $\Delta R>0.4$ 
and to have the $z$-coordinate of their tracks at the origin within $|\Delta z|<5$ cm.
The average $z$ position of any track pair must be within 4 cm of an interaction vertex ({\it primary vertex}).
Finally, the leading two electrons and muons must have tracks with an impact parameter (with respect
to the primary vertex) less than 0.02 cm, if the tracks
are reconstructed including information from the silicon detector, or less than 0.2 cm otherwise. 
The analysis is restricted to events in which a same-flavor dilepton pair with mass ($M_{ee}$ or $M_{\mu\mu}$) above 
15 GeV/$c^2$ is found;
the two highest-in-$p_T$ same-flavor leptons that satisfy this mass requirement are the leading lepton pair.
We include tau leptons that decay hadronically: they are identified as clusters of particles (jets) that have track and energy
properties expected from tau-lepton decays \cite{tau} (e.g., a decay of a tau lepton to three charged particles could result in a characteristic three-track jet substructure).
The isoTracks are not required to meet the default electron or muon requirements, but they are required 
to be isolated from other tracks, i.e., no other tracks with $p_T>0.4$ GeV/$c$ and with the same $z$ origin
as the isoTrack should be present within $\Delta R<0.4$ around the isoTrack.
Although the non-leptonic background to the isoTracks is higher, their inclusion increases the acceptance without
decreasing the sensitivity, since they are analyzed separately from
the higher-quality lepton candidates.
The isolation and topology requirements separate isoTracks and tau-lepton candidates; 
if the conditions defining both categories are satisfied, we classify the track as a tau candidate.
After the above selection, we retain 334 968 $ee$, 162 127 $\mu\mu$, 687 $ee$+$\ell$, 435 $\mu\mu$+$\ell$,
2 843 $ee$+isoTrack, and 1560 $\mu\mu$+isoTrack events.

We validate the background estimation in both two-lepton ({\it dilepton}) and trilepton final states.
The main SM dilepton background is the Drell-Yan (DY) process $q\bar{q} \rightarrow Z/\gamma^* \rightarrow \ell \ell$.  
Some electroweak background comes 
from diboson production ($WW$, $WZ$, $ZZ$, $W\gamma^*$) with subsequent leptonic
decays.  The main hadronic background contributing to the dilepton candidate sample
is the production of $W$+jets, where the $W$ boson decays to a lepton and a jet is misidentified
as a lepton (hence referred to as a {\it fake} lepton).  
Finally, top-quark-pair ($t\bar{t}$) decays that result in lepton pairs are also
included as background.
The main SM trilepton background is contributed by the production of DY dileptons in association with a photon (DY+$\gamma$), in which the photon
converts to an electron-positron pair, which, if detected, is almost always reconstructed as a single electron.
Some electroweak trilepton background come from diboson production ($WZ$, $ZZ$) with
subsequent leptonic decays.  The main hadronic background that contributes to the trilepton candidate sample
is the production of DY+jets, where a jet is misidentified as a lepton.  
Finally, $t\bar{t}$ events resulting in three leptons are also included as background.

The DY, DY+$\gamma$, diboson, and $t\bar{t}$ backgrounds are estimated with Monte Carlo (MC) simulation,
using {\sc pythia} \cite{pythia}, running with the CTEQ5L \cite{cteq} parton distribution functions,
and the CDF {\sc geant}-based \cite{geant} detector simulator.  
The MC event yields are normalized on an event-by-event basis using theoretical cross sections
(determined with next-to-leading order (NLO) quantum-field-theory calculations) \cite{cross}, event trigger efficiencies, 
lepton-identification-efficiency corrections ({\it scale factors}), and the integrated luminosity
corresponding to the CDF data sample.
\begin{figure}[!]
\hspace{-3.5cm}
\begin{minipage}[!]{2.in}
\begin{center}
\includegraphics[scale=0.44]{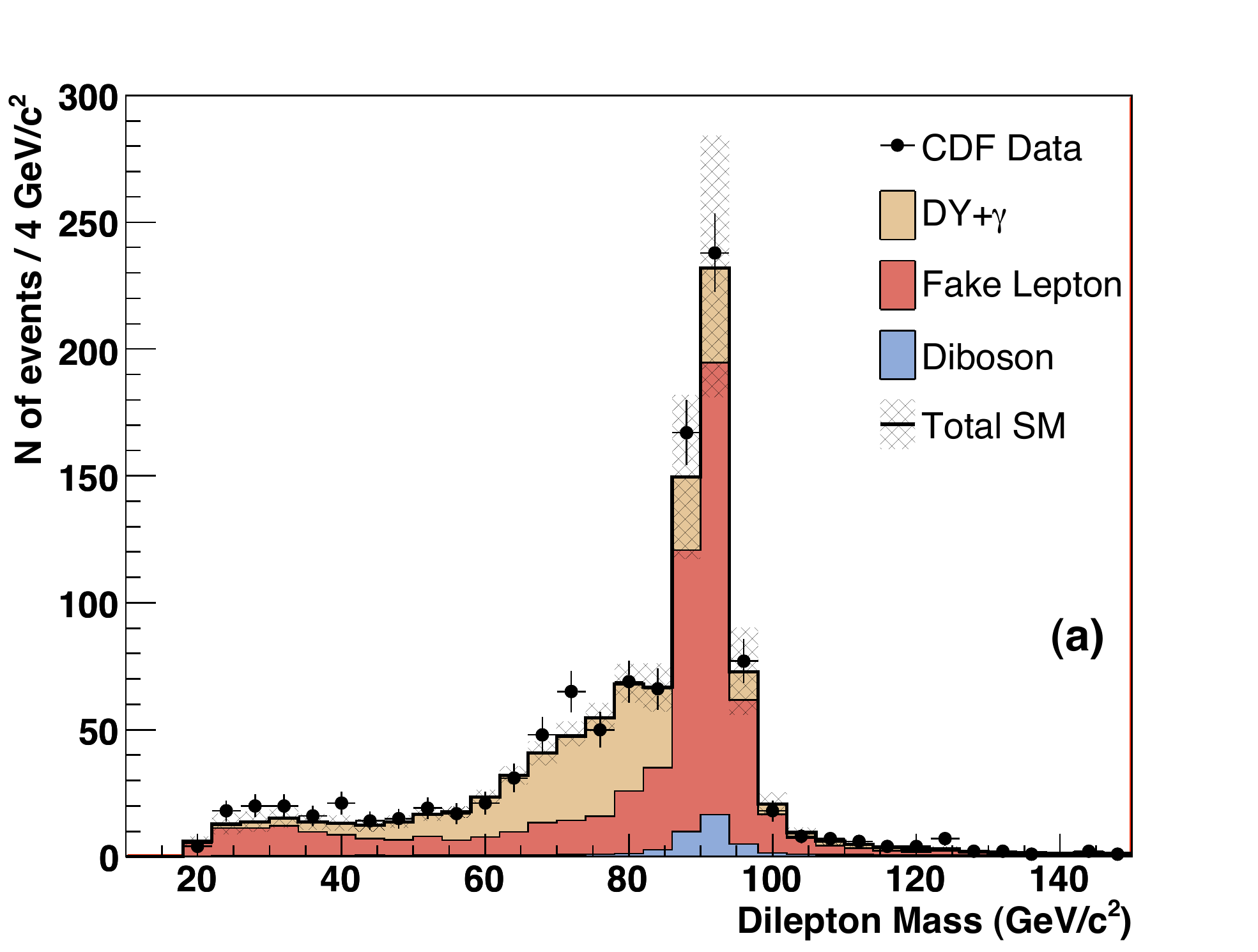}
\end{center}
\end{minipage}
\\
\hspace{-3.5cm}
\begin{minipage}[!]{2.in}
\begin{center}
\includegraphics[scale=0.44]{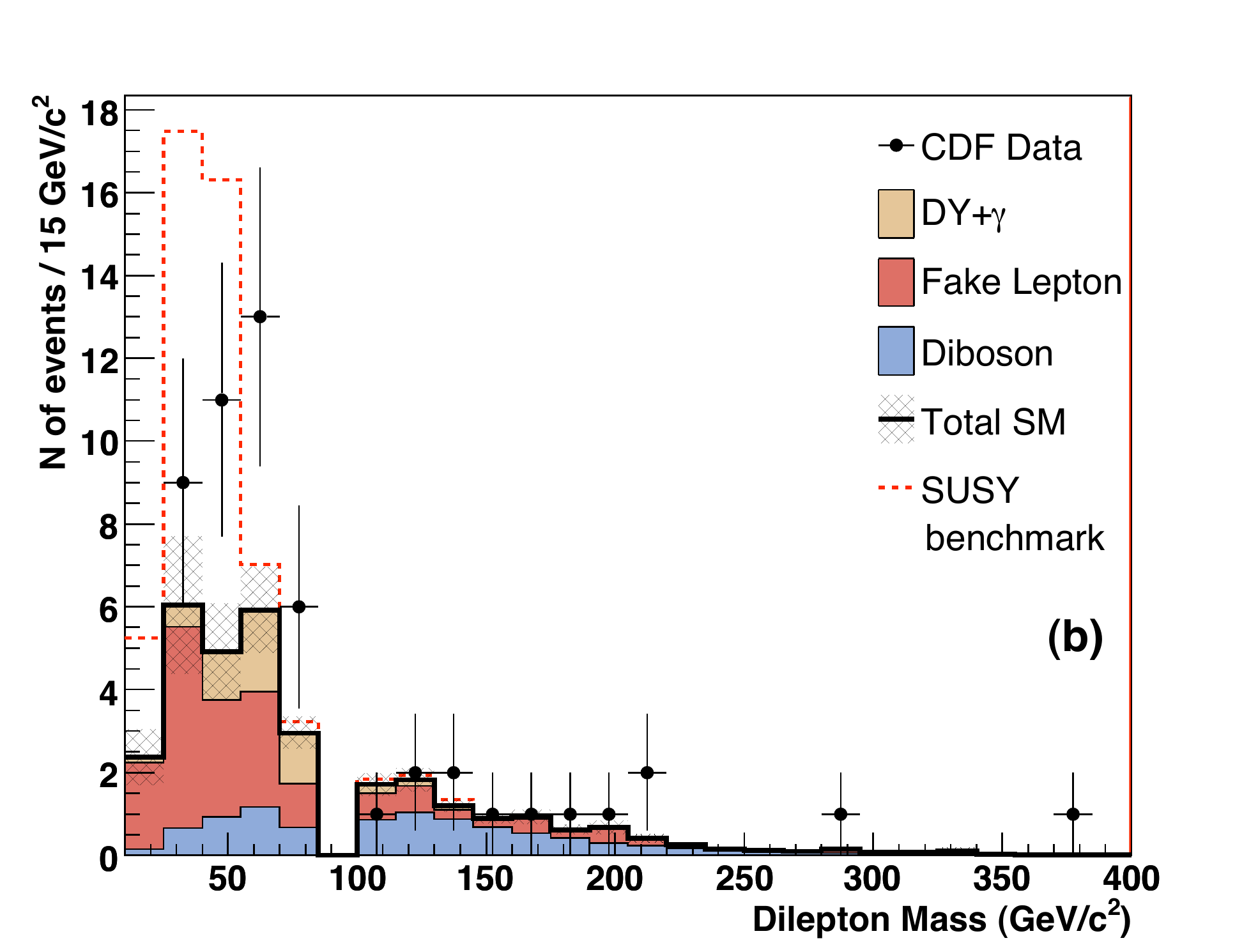}
\end{center}
\end{minipage}
\caption{(a) The leading dilepton mass distribution for the SM background (stacked histograms of DY, QCD, and diboson)
and the CDF data for $ee/\mu\mu +\ell$ events ($\ell$ is an electron, muon, or tau lepton). 
The histogram error bar shows the total SM systematic uncertainty.
(b) The dilepton mass distribution of $ee/\mu\mu +\ell$ events for the SM background, CDF data, and our mSUGRA benchmark (stacked on top of the SM background) in the signal region.  
}
\label{mass}
\end{figure}

The hadronic background in the dilepton sample, originated from quantum chromodynamic 
(QCD) processes, is estimated using CDF data, 
by selecting events with one identified lepton and applying to every well-reconstructed 
jet (track) a probability of being misidentified  as an electron (muon).
Similarly, the QCD background in the trilepton events is estimated by selecting events with 
two identified leptons of the same flavor and applying to every well-reconstructed 
jet (track) a probability of being misidentified  as an electron or as a tau lepton (muon or isoTrack).
The probabilities for a jet to be misidentified as an electron or tau lepton, or for a track to
be misidentified as a muon or isoTrack, depend on $p_T$ and on the involved detector element.  We measure the probabilities using jet-rich CDF data \cite{marcelo}.

\begin{table*}
\begin{center}
\footnotesize
\begin{tabular}{lcccccc}
\hline \hline
 {\bf Region} & {\bf Drell-Yan} & {\bf  Fakes} & {\bf  Diboson} & {\boldmath  $t\bar{t}$} & {\bf  Total SM} & {\bf Observed}\\
\hline
{\bf \boldmath A $ee$} &$1963 \pm201$&$2152 \pm524 $&$525 \pm40 $&$19 \pm2 $&{\bf  $4658 \pm583 $}&{\bf 4909}\\
{\bf \boldmath A $\mu\mu$} &$1170 \pm118$&$273 \pm136 $&$118 \pm12 $&$14 \pm1 $&{\bf  $1575 \pm181 $}&{\bf 1610}\\
{\bf \boldmath B $ee +\ell$} &$2.4 \pm0.3$&$21 \pm7 $&$13 \pm1 $&$0.13 \pm0.02 $&{\bf  $37 \pm7 $}&{\bf 35}\\
{\bf \boldmath B $\mu\mu +\ell$} &$2.6 \pm0.3$&$19 \pm6 $&$8.7 \pm0.9 $&$0.041 \pm0.01 $&{ $30 \pm6 $}&{\bf 22}\\
{\bf \boldmath B $ee$ + isoTrack} &$6.6 \pm0.7$&$249 \pm57 $&$5.8 \pm0.6 $&$0.08 \pm0.01 $&{\bf $262 \pm57 $}&{\bf 285}\\
{\bf \boldmath B $\mu\mu$ + isoTrack} &$3.3 \pm0.4$&$169 \pm38 $&$3.5 \pm0.4 $&$0.033 \pm0.009 $&{ $176 \pm38 $}&{\bf 183}\\
{\bf \boldmath C $ee +\ell$} &$86 \pm9$&$59 \pm18 $&$1.9 \pm0.2 $&$0.026 \pm0.007 $&{\bf  $147 \pm20 $}&{\bf 165}\\
{\bf \boldmath C $\mu\mu +\ell$} &$53 \pm5$&$26 \pm8 $&$0.64 \pm0.07 $&$0.015 \pm0.006 $&{ $80 \pm9 $}&{\bf 85}\\
{\bf \boldmath C $ee$ + isoTrack} &$15 \pm2$&$290 \pm58 $&$0.27 \pm0.03 $&$0.004 \pm0.003 $&{\bf $306 \pm58 $}&{\bf 270}\\
{\bf \boldmath C $\mu\mu$ + isoTrack} &$6.6 \pm0.7$&$128 \pm26 $&$0.13 \pm0.02 $&$0.004 \pm0.003 $&{ $135 \pm26 $}&{\bf 116}\\
{\bf \boldmath D $ee +\ell$} &$0.09 \pm0.02$&$0.9 \pm0.3 $&$0.12 \pm0.01 $&$0.25 \pm0.03 $&{\bf  $1.3 \pm0.3 $}&{\bf 3}\\
{\bf \boldmath D $\mu\mu +\ell$} &$0.09 \pm0.03$&$0.4 \pm0.1 $&$0.08 \pm0.01 $&$0.13 \pm0.02 $&{ $0.7 \pm0.1 $}&{\bf 0}\\
{\bf \boldmath D $ee$ + isoTrack} &$0.62 \pm0.08$&$13 \pm3 $&$0.24 \pm0.03 $&$0.65 \pm0.07 $&{\bf $14 \pm3 $}&{\bf 8}\\
{\bf \boldmath D $\mu\mu$ + isoTrack} &$0.1 \pm0.03$&$5 \pm1 $&$0.12 \pm0.01 $&$0.47 \pm0.06 $&{ $5 \pm1 $}&{\bf 2}\\
\hline
{\bf \boldmath  Signal $ee +\ell$} &$3.1 \pm0.3$&$10 \pm 4 $&$6 \pm 0.6 $&$0.44 \pm0.05 $&{\bf  $20 \pm 4 $}&{\bf 34}\\
{\bf \boldmath  Signal $\mu\mu +\ell$} &$2.6 \pm0.3$&$7 \pm2 $&$3.3 \pm0.3 $&$0.23 \pm0.03 $&{ $13 \pm2 $}&{\bf 19}\\
{\bf \boldmath  Signal $ee$ + isoTrack} & $26 \pm 3$ & $124 \pm 27$ & $6 \pm 1 $ & $0.27 \pm 0.04$ & {\bf $157 \pm 28$} & {\bf 146}\\
{\bf \boldmath  Signal $\mu\mu$ + isoTrack} &$2.8 \pm0.3$&$65 \pm15 $&$2.3 \pm0.2 $&$0.18 \pm0.03 $&{ $70 \pm15 $}&{\bf 62}\\
\hline \hline
\end{tabular}
\caption{Expected and observed event yields in the main control regions (A, B, C, D, as described in the text)
that are used to confirm the SM background estimation and in the signal region.
Here $\ell$ is an electron, muon, or tau lepton.  The DY background for trileptons includes a photon, which converts and is reconstructed as an electron.  All uncertainties include systematic contributions only.}
\label{t_mm}
\end{center}
\end{table*}

The main sources of systematic uncertainty on the MC-estimated backgrounds \cite{syst}
are the theoretical cross sections (an 8\% effect on the event yields),
the luminosity (6\%), the lepton-ID efficiency (2\%), the parton distribution functions (2\%), 
and the trigger efficiency (0.5\%).
The total systematic uncertainty on the expected event yield is $\sim$10\%.  
The QCD background systematic uncertainty is $\sim$50\% for falsely identified electrons and muons
with transverse momentum greater than 20 GeV/$c$ and $\sim$20\% for lower transverse momentum.
This uncertainty is estimated from the variation in the measurement of the misidentification rates using 
different jet-rich CDF data sets triggered with varied jet-energy thresholds.  

In order to validate our background estimates, we investigate dilepton and trilepton 
control regions defined by restricting events to specific regions of the multidimensional space defined by the leading dilepton mass $M_{ee/\mu\mu}$, 
the missing transverse momentum $\met$ \cite{footnote}, and the jet multiplicity $N_j$.
For an unbiased selection of events, we avoid looking at the data in the signal region, which is defined
as trilepton events with ($15<M_{ee/\mu\mu}<76$ GeV/$c^2$ or $M_{ee/\mu\mu}>106$ GeV/$c^2$), 
$\met>15$ GeV, and $N_j\le 1$.  We define the control regions by inverting at least one of the signal-region
selection requirements.  Overall, 24 dilepton and 40 trilepton control regions are used.  
One of the most critical control regions consists of
dilepton events selected as signal but without requiring a third lepton (region A); the trilepton signal region is 
a subset of region A.  We also present here trilepton control regions with only one of the three signal-region requirements inverted: 
either dilepton mass in the $Z$ boson resonance ($76<M_{ee/\mu\mu}<106$ GeV/$c^2$), 
or $\met<10$ GeV, or $N_j\ge 2$, which lead to regions B, C, and D respectively. 
Region A is used to validate all sources of background in the dilepton signal region.  Region B is used to validate the diboson background estimates; 
region C, the DY and fake-lepton backgrounds; and region D, the top-quark background, all in the trilepton subset of the data.
The QCD background estimation is validated in the intermediate-mass ($20<M_{ee/\mu\mu}<76$~GeV/$c^2$) control region,
as well as in the trilepton ($76<M_{ee/\mu\mu}<106$~GeV/$c^2$) and high-mass $t\bar{t}$ control regions.
The $t\bar{t}$ dilepton background is also validated in a control region consisting of events with
two or more hadronic jets, $H_T>200 $~GeV \cite{footnote2}, and $\met>20$~GeV, where the top-quark pair 
production is the dominant process.  Finally, good agreement between SM expectation and $Z$-resonance data supports the estimation 
of efficiencies, scale factors, data-set luminosity and theoretical cross sections.  

Table I shows the expected and observed event yields in these control regions, where good agreement is observed.  
The same is true for all other control regions \cite{trilep_public}.
Overall, we observe 260 010 dielectrons 
and 142 386 dimuons in the $Z$-resonance region, where we expect $268$ $670 \pm 26$ $486$ and $146$ $103 \pm 14$ $573$ respectively 
(systematic uncertainties only).  Figure 1(a) shows the leading dilepton mass distribution for the observed $ee/\mu\mu+\ell$ 
events, along with the SM expectation.

After observing satisfactory agreement between SM expectation and experimental observation in all the control regions, we uncover the data in the signal region.  We observe 34 $ee$+$\ell$, 146 $ee$+isoTrack, 19 $\mu\mu$+$\ell$, and 62 $\mu\mu$+isoTrack events, whereas the SM expectations are $20 \pm 4$, $157 \pm 28$, $13 \pm 2$, and $70 \pm 15$ respectively (systematic uncertainties only).  Figure 1(b) shows the leading dilepton mass distribution for $ee$+$\ell$ and $\mu\mu$+$\ell$ events in the signal region for SM background, our mSUGRA benchmark point \cite{benchmark} ($m_0=60$ GeV/$c^2$, $m_{1/2}=190$ GeV/$c^2$, $\tan\beta=3$, $A_0=0$, and $\mu>0$), and observation.  
A moderate excess of events is observed in the four dilepton mass bins between 30 and 80 GeV/$c^2$, 
whose significance is estimated as follows.
The probability that an excess of the same or larger size is seen within four consecutive bins (range of 60 GeV/$c^2$) anywhere in the dilepton mass spectrum of $e/\mu+\ell$, assuming no new physics, corresponds to a $p$-value of  0.032 (1.85$\sigma$).  This probability is determined with the use of pseudoexperiments that take into account the statistical and systematic uncertainties of the actual experiment.  In the fakes-dominated $e/\mu$+isoTrack signal-region, results are more consistent with the SM ($p$-value = 0.56).

These results are used to set limits on the associated
chargino-neutralino production rates and exclude part of the ($m_{\tilde{\chi}_2^0}$ vs. $m_{\tilde{\tau}}$) space,
which is investigated with an mSUGRA parameter scan that varies $m_{0}$ and $m_{1/2}$ and fixes the other parameters at the benchmark values.
For the chargino-neutralino upper cross-section limits, we simulate SUSY events 
with corresponding gaugino masses $m_{\bar{\chi}_1^{\pm}} =97-200$ GeV/$c^2$ and $m_{\bar{\chi}_1^0}=55-108$ GeV/$c^2$.  
The SUSY MC events are produced and normalized in the same manner 
as the background-MC events and are characterized by the same sources of systematic uncertainty.  
The SUSY mass spectrum is calculated 
using {\sc softsusy} \cite{softsusy} and the next-to-leading-order chargino-neutralino production
cross section is determined using {\sc prospino} \cite{prospino}.  Finally the {\sc isasusy} function of
{\sc isajet} \cite{isajet} is used to determine the branching ratios of charginos and neutralinos to leptons.
The CDF acceptance for the trilepton SUSY signal is $\sim$2\%.

\begin{figure}[t]
\includegraphics[scale=0.44]{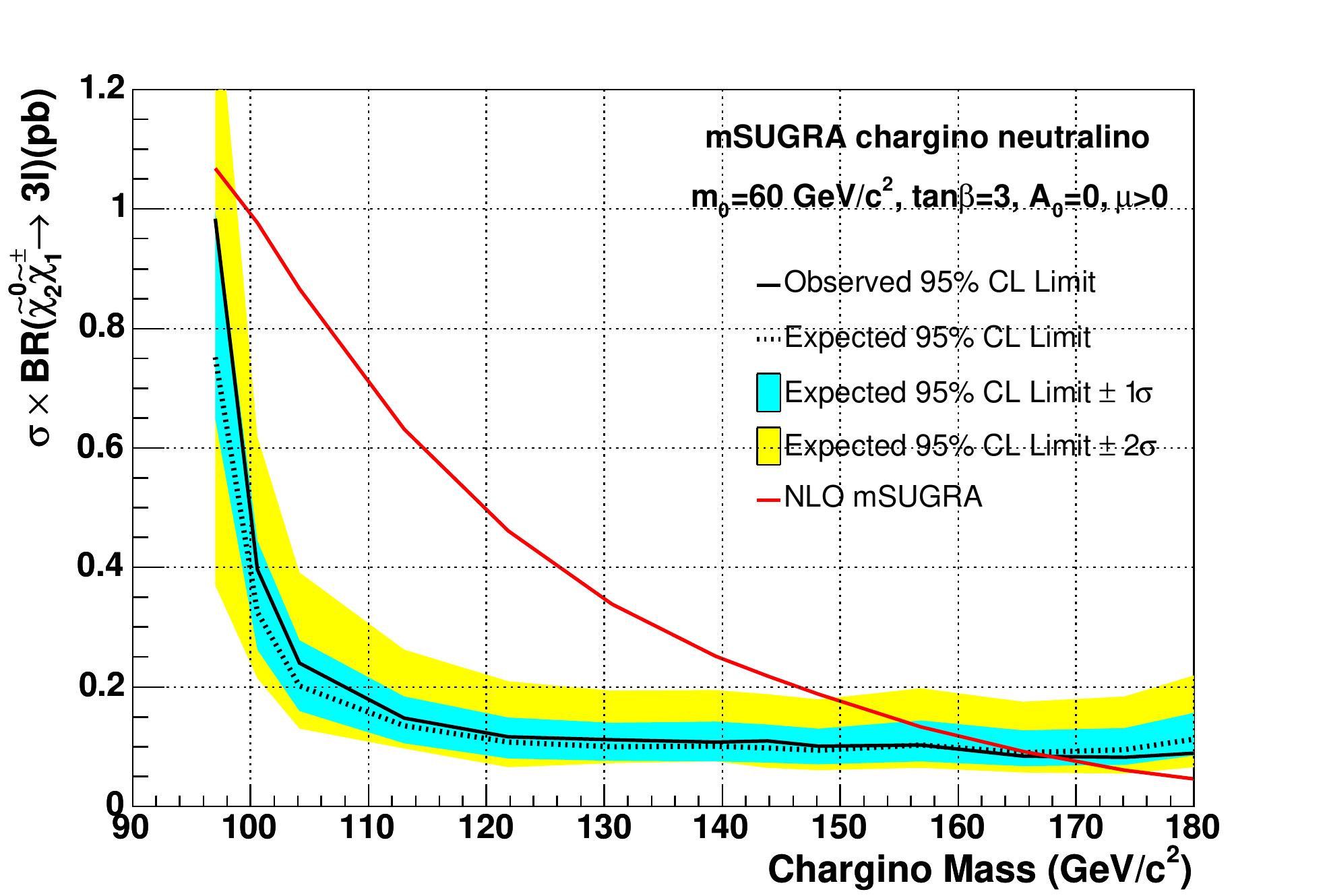}
\caption{The 95\% C.L. upper limit on associated chargino-neutralino production cross section as a function of the chargino's mass.
The region above the limit lines are excluded.  Intersections with the theoretical NLO mSUGRA prediction give the 95\% C.L.
chargino mass limits.}
\label{limit1}
\end{figure}

To increase sensitivity to a SUSY signal, we optimize the selection separately 
for each mSUGRA spectrum point, using the ratio between the SUSY-signal strength and the uncertainty on the SM-background prediction as figure of merit.
In the optimization process we treat all trilepton channels separately.  The resulting optimal requirements include the $\met>25$ GeV criterion and the kinematic constraint $M_{ee/\mu\mu}<m_{\tilde{\chi}^{\pm}_1}-m_{\tilde{\chi}^{0}_1}$.  We also optimize the transverse-momentum requirement for the three leptons as well as the subleading-dilepton-mass requirement \cite{trilep_public}.  The limits are set using a modified frequentist method approach ($CL_{s}$ method) \cite{tomjunk,read} that compares the background-only with the signal-plus-background hypotheses, treating all trilepton channels independently.  
Figure 2 shows the 95\% confidence level ($\cl$) cross-section ($\sigma \times {\rm BR}(\tilde{\chi}^{\pm}_1\tilde{\chi}^{0}_2\rightarrow \ell \ell \ell$)) 
exclusion upper limit as a function of the lightest chargino mass $m_{\tilde{\chi}_1^{\pm}}$, along with the theoretical cross-section.  
The intersection of the cross-section exclusion limit  with the theoretical cross-section curves gives the 95\% $\cl$ lower limit on $m_{\tilde{\chi}_1^{\pm}}$.  Masses above 96 GeV/$c^2$ and below 168 GeV/$c^2$ are excluded.  For $140<m_{\tilde{\chi}_1^{\pm}}<180$~GeV/$c^2$, the trilepton analysis excludes cross sections greater than 0.1 pb at the 95\% $\cll$.  
\begin{figure}[t]
\includegraphics[scale=0.44]{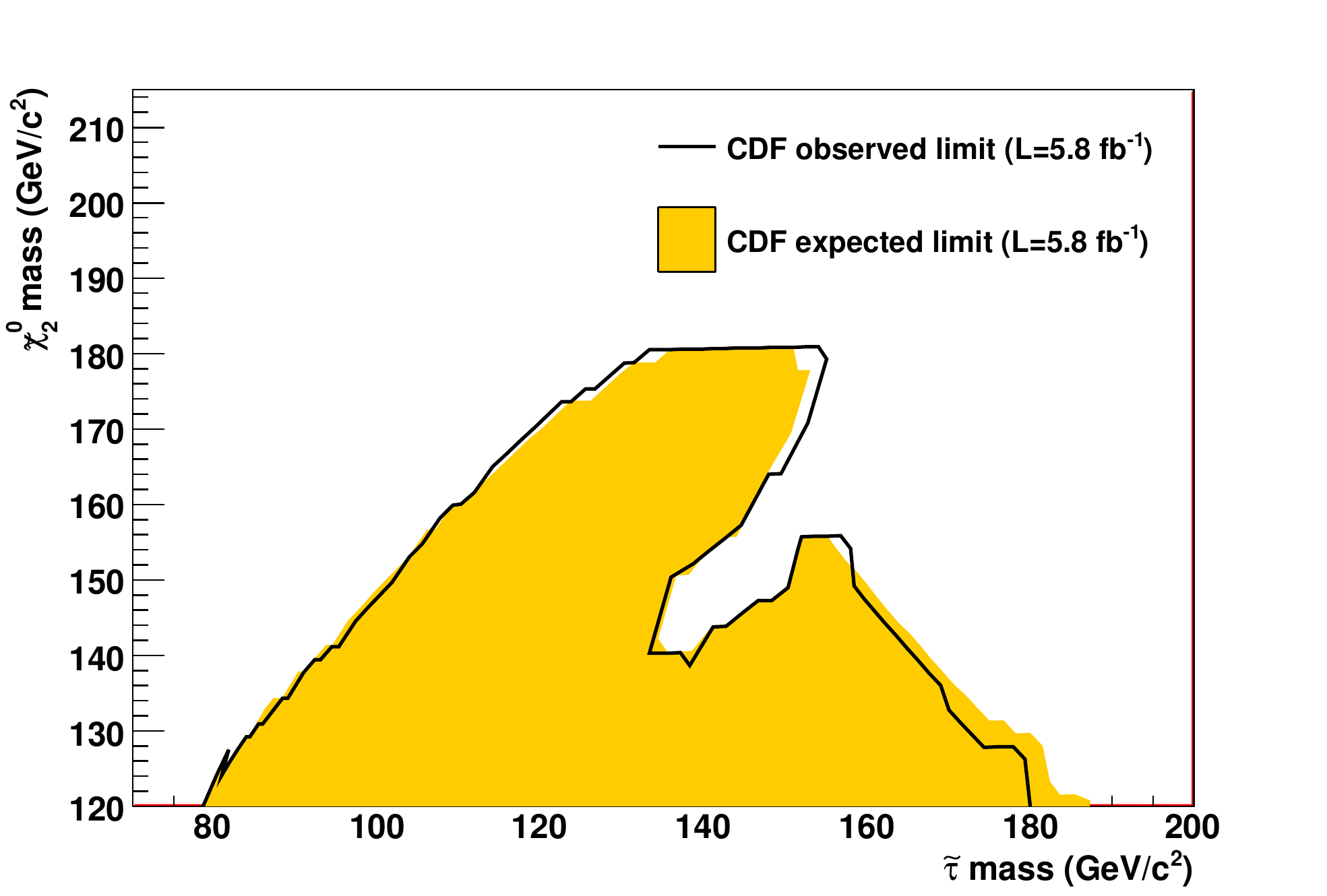}
\caption{The 95\% C.L. exclusion in the ($m_{\tilde{\chi}_2^0}$ vs. $m_{\tilde{\tau}}$) space.}
\label{limit2}
\end{figure}

We repeat the procedure by varying the masses of the next-to-lightest neutralino $\tilde{\chi}_2^0$ and $\tilde{\tau}$ and report the corresponding two-dimensional exclusion region shown in Fig. 3.
This analysis excludes part of the ($m_{\tilde{\chi}_2^0}$ vs. $\tilde{\tau}$) space not 
excluded in previous CDF or D0 results (\cite{ucsd,d0}) due to its additional sensitivity to decays 
of tau leptons into hadrons and low-$p_T$ leptons.  We are sensitive to mass differences $m_{\tilde{\chi}_2^0} - m_{\tilde{\tau}} \gtrsim$ 15 GeV/$c^2$.

In summary, we present a search for new physics in the trilepton+$\met$ final state using 
data from 1.96-TeV proton-antiproton collisions collected by CDF and
corresponding to an integrated luminosity of 5.8 fb$^{-1}$.
In the study, we include low-momentum leptons that are not investigated at the LHC
and that could result from direct decays of new light particles or chain decays of particles 
with similar masses.
We do not observe any significant discrepancies from the expected SM prediction. 
We exclude cross sections up to $0.1$ pb and chargino masses up to $168$ GeV/$c^2$
at 95\% $\cl$ for the mSUGRA parameters $m_0=60$ GeV/$c^2$, $\tan\beta=3$, and $A_0=0$
, and establish an exclusion in the ($m_{\tilde{\chi}_2^0}$ vs. $m_{\tilde{\tau}}$) space.

We thank the Fermilab staff and the technical staffs of the
participating institutions for their vital contributions. This work
was supported by the U.S. Department of Energy and National Science
Foundation; the Italian Istituto Nazionale di Fisica Nucleare; the
Ministry of Education, Culture, Sports, Science and Technology of
Japan; the Natural Sciences and Engineering Research Council of
Canada; the National Science Council of the Republic of China; the
Swiss National Science Foundation; the A.P. Sloan Foundation; the
Bundesministerium f\"ur Bildung und Forschung, Germany; the Korean
World Class University Program, the National Research Foundation of
Korea; the Science and Technology Facilities Council and the Royal
Society, United Kingdom; the Russian Foundation for Basic Research;
the Ministerio de Ciencia e Innovaci\'{o}n, and Programa
Consolider-Ingenio 2010, Spain; the Slovak R\&D Agency; the Academy
of Finland; the Australian Research Council (ARC); and the EU community
Marie Curie Fellowship Contract No. 302103.


\begin{thebibliography}{99}
\bibitem[{fooo({\natexlab{a}})}]{msugra}
\bibinfo{note}{In mSUGRA (minimal supergravity) the only free parameters of the theory are the common scalar mass 
$m_0$, the common gaugino mass $m_{1/2}$, the ratio of Higgs vacuum expectation values tan$\beta$, 
the trilinear sfermion-sfermion-Higgs coupling $A_0$, and the sign of the higgsino scale parameter $\mu$.}
\bibitem[{foo({\natexlab{a}})}]{footnote}
\bibinfo{note}{The missing transverse momentum vector $\vec{\met}$ is defined as 
$-(\sum_i \vec{E}_{i})_T$, where $\vec{E}_{i}$ has magnitude equal to the energy deposited in the 
i-th calorimeter tower and direction perpendicular to the beam axis and pointing to that calorimeter 
tower at $\eta=0$.  The $\met$ is corrected for the presence of muons, because they deposit only a little of their
energy in the calorimeters.}
\bibitem[{foo({\natexlab{a}})}]{footnote2}
\bibinfo{note}{$H_T$ is defined as the scalar sum of the transverse momenta 
of jets, leptons and $\met$}.
\bibitem{benchmark}
\bibinfo{note}{Our mSUGRA benchmark point is an indicative point in the 
mSUGRA parameter space, for which $m_{\tilde{\chi}_1^{\pm}}\approx m_{\tilde{\chi}_2^0}\approx 123$ GeV/$c^2$,
and which leads to decays to three leptons, with a preference to lower momenta.  For this point the 
allowed decays of gauginos are
$\tilde{\chi}^{\pm}_1 \rightarrow \tilde{\tau}\nu_{\tau}\rightarrow \ell \nu \tilde{\chi}^{0}_1$ and 
$\tilde{\chi}^{0}_2 \rightarrow \tilde{\ell} \ell \rightarrow \ell \ell \tilde{\chi}^{0}_1$}.
\end{thebibliography}
\end{document}